\documentclass{article}
\usepackage{graphicx} 
\usepackage{authblk}
\usepackage{hyperref}
\usepackage[
backend=biber,
style=nature,
sorting=none
]{biblatex}
\title{SCIMAP: A Python Toolkit for Integrated Spatial Analysis of Multiplexed Imaging Data}
\author[1,2,3]{Ajit J. Nirmal\thanks{Corresponding author: \texttt{anirmal@bwh.harvard.edu}}}
\author[2,3]{Peter K. Sorger}
\affil[1]{Department of Dermatology, Brigham and Women’s Hospital, Harvard Medical School, Boston, MA, United States of America}
\affil[2]{Ludwig Center for Cancer Research at Harvard, Harvard Medical School, Boston, MA, United States of America}
\affil[3]{Laboratory of Systems Pharmacology, Harvard Medical School, Boston, MA, United States of America}

\date{}
\begin{document}
\maketitle
\section{Summary}

Multiplexed imaging data are revolutionizing our understanding of the composition and organization of tissues and tumors \cite{noauthor_catching_2022}. A critical aspect of such “tissue profiling” is quantifying the spatial relationships among cells at different scales from the interaction of neighboring cells to recurrent communities of cells of multiple types. This often involves statistical analysis of 10$^7$ or more cells in which up to 100 biomolecules (commonly proteins) have been measured. While software tools currently cater to the analysis of spatial transcriptomics data \cite{liu_analysis_2022}, there remains a need for toolkits explicitly tailored to the complexities of multiplexed imaging data including the need to seamlessly integrate image visualization with data analysis and exploration. We introduce SCIMAP, a Python package specifically crafted to address these challenges. With SCIMAP, users can efficiently preprocess, analyze, and visualize large datasets, facilitating the exploration of spatial relationships and their statistical significance. SCIMAP's modular design enables the integration of new algorithms, enhancing its capabilities for spatial analysis.

\section{Statement of Need}

A variety of methods have been introduced for high multiplexed imaging of tissues, including MxIF, CyCIF, CODEX, 4i, mIHC, MIBI, IBEX, and IMC \cite{angelo_multiplexed_2014, gerdes_highly_2013, giesen_highly_2014, goltsev_deep_2018, gut_multiplexed_2018, tsujikawa_quantitative_2017, lin_highly_2018}; although these methods differ in their implementations, all enable the collection of single-cell data on 20-100 proteins within preserved 2D and 3D tissue microenvironments. Analysis of high-plex images typically involves joining adjacent image tiles together and aligning channels across imaging cycles (stitching and registration) to generate a composite high-plex image and then identifying the positions and boundaries of individual cells via segmentation. The intensities of individual protein antigens, stains, and other detectable molecules are then quantified on a per-cell basis. This generates a “spatial feature table” (analogous to a count table in sequencing) that can be used to identify individual cell types and states; tools from spatial statistics are then used to identify how these cells are patterned in space from scales ranging from a few cell diameters (~10 µm) to several millimeters.

Spatial feature tables provide the quantitative data for analysis of high-plex data but human inspection of the original image data remains essential. At the current state of the art, many of the critical morphological details in high-resolution images cannot be fully and accurately quantified. Segmentation is also subject to errors identifiable by humans, but not fully resolvable computationally \cite{baker_quality_2024}. As a consequence, computation of spatial features and relationships must be performed in combination with visualization of the underlying image data. Humans excel at identifying tissue features that correspond to classical histo-morphologies; they are also effective at discriminating foreground signals from variable background \cite{nirmal_cell_2023} using a process of “visual gating” (perception of high and low-intensity levels while visualizing an image). More generally, effective integration of visualization and computation enables nuanced interpretation of cellular organization in relation to established tissue architectures. SCIMAP uses the Python-based Napari \cite{chiu_napari_2022} image viewer to leverage these capabilities by providing a seamless interface to inspect and annotate high-plex imaging data alongside computational analysis. For example, we have implemented an image-based gating approach that allows users to visually determine the threshold that discriminates background from a true signal at both a whole-specimen and single-cell level. Users can also select specific regions of interest (ROIs) for selective or deeper analysis. This involves drawing ROIs over images (freehand or geometric) and then selecting the underlying single data for further analysis. This capability is essential for incorporating histopathological information on common tissue structures (e.g., epidermis, dermis, follicles), immune structures (e.g., secondary and tertiary lymphoid structures), tumor domains (e.g., tumor center, boundary, tumor buds), and tumor grade or stage (e.g., early lesions, invasive regions, established nodules). It also allows for excluding regions affected by significant tissue loss, folding, or artifactual staining \cite{baker_quality_2024}. SCIMAP then performs statistical and spatial analyses on individual ROIs or sets of ROIs. Spatial analysis, including the measurement of distances between cells, analysis of interaction patterns, categorization into neighborhoods, and scoring of these patterns, is crucial for elucidating the cellular communications that underpin the functional aspects of the biology being studied. SCIMAP offers various functions to facilitate these analyses. 

Lastly, a single high-plex whole slide image can exceed 100GB per image and 10$^6$ cells, necessitating optimized functions for handling large matrices and images. SCIMAP employs the well-established AnnData object structure, complemented by Dask and Zarr for efficient image loading in Napari. This approach facilitates seamless viewing of images with overlaid data layers, thus enabling effective analysis of large datasets. To date, SCIMAP has been used in the analysis of over 5 datasets from 8 tissue and cancer types \cite{yapp_multiplexed_2024, nirmal_spatial_2022, gaglia_lymphocyte_2023, maliga_immune_2024}.

\section{Availability and Features}

\begin{figure}[h]
    \centering
    \includegraphics[width=1\linewidth]{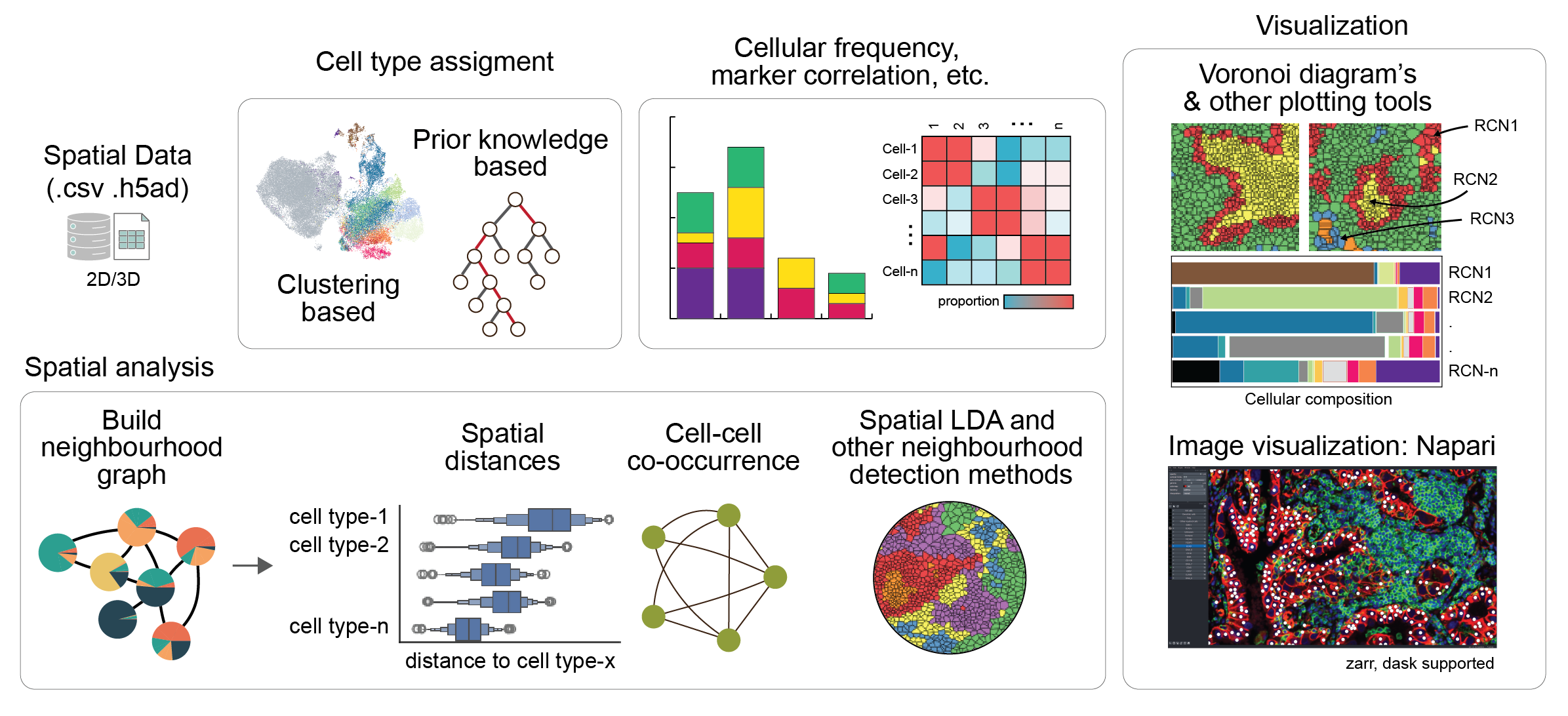}
    \caption{SCIMAP Workflow Overview. The schematic highlights data import, cell classification, spatial analysis, and visualization techniques within the SCIMAP tool box.}
    \label{fig:workflow}
\end{figure}

SCIMAP is available as a standalone Python package for interactive use, in Jupyter Notebook for example, or can be accessed via a command-line interface (CLI; only a subset of functions that do not require visualization) for cloud-based processing. The package can be accessed at GitHub (\url{https://github.com/labsyspharm/scimap}) and installed locally through pip. Installation, usage instructions, general documentation, and tutorials, are available at \url{https://scimap.xyz/}. See Figure ~\ref{fig:workflow} for a schematic of the workflow and system components.

SCIMAP comprises of four main modules: preprocessing, analysis tools, visualization, and external methods. The preprocessing tools include functions for normalization, batch correction, and streamlined import from cloud processing pipelines such as MCMICRO \cite{schapiro_mcmicro_2022}. The analysis tools offer standard single-cell analysis techniques such as dimensionality reduction, clustering, prior knowledge-based cell phenotyping (a method through which cells are classified into specific cell types based on patterns of marker expression defined by the user), and various spatial analysis tools for measuring cellular distances, identifying regions of specific cell type aggregation, and assessing statistical differences in proximity scores or interaction frequencies. SCIMAP also includes neighborhood detection algorithms that utilize spatial-LDA \cite{wang_spatial_2007} for categorical data (cell types or clusters) and spatial lag for continuous data (marker expression values). Most analysis tools come with corresponding visualization functions to plot the results effectively. Additionally, the external methods module facilitates the integration of new tools developed by the community into SCIMAP, further extending its utility and applicability to both 2D and 3D data. 

\section{Acknowledgements}

This work was supported by NCI grant R00CA256497 to A.J.N, U01-CA284207 to P.K.S and by Ludwig Cancer Center at Harvard. We would like to thank our many users and members of the Laboratory for Systems Pharmacology at Harvard Medical School for their invaluable feedback.

\printbibliography
\end{document}